\documentclass[journal]{IEEEtran}
\usepackage{amsmath,amsfonts,amssymb}
\usepackage{algorithmic}
\usepackage{algorithm}
\usepackage{array}
\usepackage[caption=false,font=normalsize,labelfont=sf,textfont=sf]{subfig}
\usepackage{textcomp}
\usepackage{stfloats}
\usepackage{url}
\usepackage{verbatim}
\usepackage{graphicx}
\usepackage[utf8]{inputenc}
\usepackage[citestyle=numeric-comp,sorting=none]{biblatex}
\usepackage{setspace}
\usepackage{multirow}

\usepackage{lipsum,amsmath,verbatim}
\usepackage{siunitx}
\usepackage{cleveref}
\usepackage{tikz}
\usepackage{flushend}
\usepackage{xcolor}
\usepackage[inline]{enumitem}

\AtNextBibliography{\footnotesize}

\usetikzlibrary{shapes.misc}
\usetikzlibrary{shapes.geometric}
\usetikzlibrary{calc}

\tikzset{cross/.style={cross out, draw=black, minimum size=2*(#1-\pgflinewidth), inner sep=0pt, outer sep=0pt},
cross/.default={2pt}}

\newif\ifChanges
\Changesfalse

\ifChanges
\newcommand{\changed}[1]{{\color{orange}{#1}}}%
\newcommand{\removed}[1]{\sout{#1}}%

\usepackage[%
linewidth=1pt,
middlelinecolor= orange,
linecolor=orange,
middlelinewidth=0.4pt,
roundcorner=1pt,
topline = false,
leftline = false,
rightline = true,
bottomline = false,
rightmargin=-.1cm,
innerleftmargin=0cm,
innerrightmargin=.1cm,
innertopmargin=0pt,
innerbottommargin=0pt,
]{mdframed}

\newmdenv[
middlelinecolor= green,
linecolor=green,
]{newcontent}

\else 

\newcommand{\changed}[1]{#1}%
\newcommand{\removed}[1]{}

\fi

\begin{document}

\title{Privacy Disclosure of Similarity Rank \\ in Speech and Language Processing}

\author{Tom Bäckström,~\IEEEmembership{Senior Member,~IEEE,} Mohammad Hassan Vali,~\IEEEmembership{Member,~IEEE}, 
My Nguyen, Silas Rech
\thanks{Bäckström, Nguyen and Rech are with Aalto University, Department of Information and Communications Engineering, Finland, email:first.lastname@aalto.fi. Vali is with Aalto University, Department of Computer Science, Finland. }
\thanks{Manuscript received April 19, 2021; revised August 16, 2021.}%
\thanks{A part of this work was supported by ‘‘Designing Inclusive \&
Trustworthy Digital Public Services for Migrants in Finland (Trust-M)’’
(Grant Number: 353529) funded by the Strategic Research Council of
Finland.}}

\markboth{IEEE Transactions on Audio Speech and Language Processing,~Vol.~XX, No.~Y, Month~202Z}%
{Bäckström \MakeLowercase{\textit{et al.}}: Privacy of Streamed Speech Data}

\IEEEpubid{0000--0000/00\$00.00~\copyright~2021 IEEE}

\maketitle

\begin{abstract}
Speaker, author, and other biometric identification applications often compare a sample's similarity to a database of templates to determine the identity. Given that data may be noisy and similarity measures can be inaccurate, such a comparison may not reliably identify the true identity as the most similar. Still, even the similarity rank based on an inaccurate similarity measure can disclose private information about the true identity. We propose a methodology for quantifying the privacy disclosure of such a similarity rank by estimating its probability distribution. It is based on either determining the histogram of the similarity rank for the true speaker or, when data are scarce, modeling the histogram with the beta-binomial distribution. We express the disclosure in terms of entropy (bits), such that the disclosure from independent features is additive. Our experiments demonstrate that all tested speaker and author characterizations contain personally identifying information (PII) that can aid in identification, with embeddings from speaker recognition algorithms containing the most information, followed by phone embeddings, linguistic embeddings, and fundamental frequency. Our initial experiments show that the disclosure of PII increases with the length of test samples, but it is bounded by the length of database templates. The provided metric, similarity rank disclosure, provides a way to compare the disclosure of PII between biometric features and merge them to aid identification. It can thus aid in the holistic evaluation of threats to privacy in speech and other biometric technologies.

\end{abstract}

\section{Introduction}
Speech is a tool for communication that conveys private information from one agent to another. In addition to the legitimate message, speech contains plenty of bundled side information, such as state of health, emotions, identity, and background, most of which is private~\cite{singh2019profiling,singh2025humanvoiceunique}. Speech technology, which processes, transmits, or stores speech, can thus expose users to a multitude of threats, such as price gouging, stalking, and identity theft~\cite{backstrom2023privacy}. Protecting user privacy in speech applications is morally important, but since threats to users also reduce user satisfaction, it is also important for business. 

To accurately characterize privacy threats, we need to quantify them. In speech technology, the predominant approach for identifying speakers is to compare an input signal to a database of speaker templates, and the speaker in the database most similar to the input is chosen as the identified user. However, if the database is large or if the speaker identification approach is noisy, we are unlikely to identify the correct speaker. Still, if a particular speaker template is a ``relatively good'' match to the observation, it does support the hypothesis that this template is the true identity.  In other words, even when the true speaker is not the most similar in the database, observations can still disclose some private information. 
\IEEEpubidadjcol

Categories of speakers' properties that can be used to identify them are plentiful and include at least physical and psychological characteristics and state,  linguistic content and style, language used, proficiency and accent, affiliations, relationship character, acoustic environment, as well as hardware and software used for recording~\cite{backstrom2023privacy,singh2019profiling}. Information in every such category can potentially aid in identifying a speaker, in particular if an attacker has access to multiple categories of information~\cite{singh2025humanvoiceunique}. \changed{It is thus essential to quantify privacy in a way that enables joint measurement across combinations of categories.}

\begin{figure}[t!]
    \centering
    \includegraphics[width=\columnwidth]{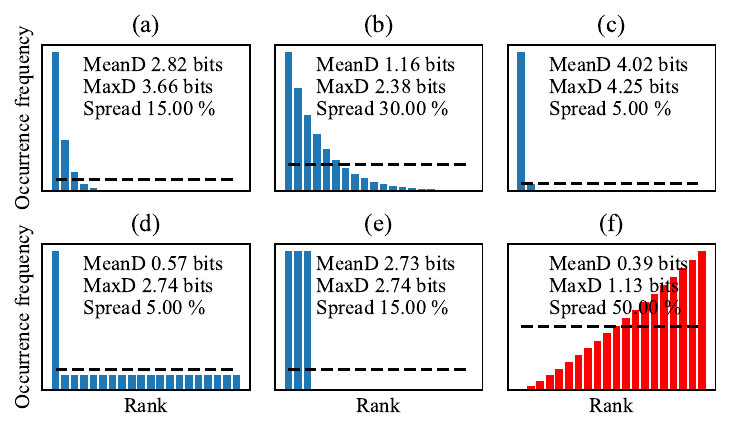}
    \caption{Illustrations of similarity rank distribution (a) with a typical distribution, (b) with a wide distribution corresponding to noisy data or a poor similarity metric, (c) a narrow distribution corresponding to a good similarity metric and data, (d) where non-matching templates have equal probability, (e) where the most similar templates have equal probability and (f) an anomalous distribution where higher ranks have higher probability. The black dashed line indicates the level of equal probability for all ranks. The mean and maximum disclosures (MeanD and MaxD, respectively) and spread of ranks is listed in each pane (see \cref{sec:stats}).}
    \label{fig:samplehist}
\end{figure}

\changed{The purpose of this paper is to quantify the privacy disclosure of observed input data by using rank-ordered similarity to a set of reference samples and express it as the reduction in entropy obtained through the observation.} It applies to scenarios where similarity measures are imperfect or where data is noisy (see \cref{sec:threats}). We provide a methodology for quantifying such disclosure from histograms of similarity rank extracted from large databases (see \cref{sec:methods}). We also derive a model that can be applied to small or sparse datasets, based on the beta-binomial distribution. We apply the methodology and model to histograms of fundamental frequencies and phone distributions, speaker embeddings from a speaker recognition algorithm, and linguistic embeddings (see \cref{sec:features}) using the Fisher and VoxCeleb corpuses (see \cref{sec:data}). Our experiments in \cref{sec:experiments} demonstrate that the beta-binomial distribution is an accurate model of empirical distributions of similarity rank.
Additionally, we find that speech attributes carry identifying information (in an increasing order) in fundamental frequency distribution, linguistic style, phone distribution, and speaker identity embeddings. The magnitude of disclosure depends on both the length of the observation (test sample) and the template database entries. The disclosure, however, saturates after a certain length of the test/template, depending on the length of the other.

\section{Background}
The predominant approach for privacy-preserving processing of speech signals is \emph{disentanglement}, where the signal is decomposed into components representing separate features of the speech signal, such as its phonetic content, intonation, and speaker identity~\cite{pierre22_interspeech,backstrom2023privacy,li21e_interspeech,sun2016phonetic}. Provided that the disentangled channels represent independent information, they can be selectively anonymized through replacement, distortion, or quantization~\cite{pierre22_interspeech,vali2024porcupine}. Replacement of speaker characteristics is known as \emph{voice conversion}, and when the purpose of such conversion is to protect privacy, it is known as \emph{anonymization}.

Though disentanglement is an essential element of privacy-preserving speech processing, it is crucial to understand where private information resides and who has access to it. A recommended approach to studying, characterizing, and encapsulating such information is a scenario of use or threat model~\cite{rahman24_spsc,backstrom2023privacy}. In particular, in scenarios involving interactions between local (edge) devices and cloud servers, it may be beneficial to protect privacy as early in the pipeline as possible, or in practice, on the edge device~\cite{wu2021understanding,backstrom2023privacy}. This limits the opportunities that stakeholders and attackers have to abuse private information.

A notable activity in privacy-preserving speech research is the semi-regular VoicePrivacy Challenge~\cite{Panariello2024voiceprivacy}, where privacy-preserving processing methods are evaluated.
The natural counterpart to privacy-preserving speech processing is speaker recognition~\cite{BAI202165,vaessen2022fine}, where the objective is to identify \emph{who} is speaking. Typically, both privacy-preserving algorithms and speaker recognition are evaluated with metrics related to the Equal Error Rate (EER), where the detection threshold is adjusted such that false acceptance rate (FAR) and false rejection rate (FRR) are equal~\cite{brown2022voxsrc2021voxcelebspeaker,Panariello2024voiceprivacy}. These measures, however, have significant drawbacks: they quantify only the mean exposure, whereas worst-case exposure should also be considered~\cite{nautsch20_interspeech}; they employ pairwise comparison without an explicit joint probability model, and they do not address the magnitude of disclosure for private information.

A significant proportion of research on privacy is related to databases, where fixed-length data is communicated between agents. This work, in contrast, applies also to streaming data, whose impact on privacy is by nature more difficult to analyze~\cite{ali2005hippocratic}. A series of queries to the same database share similarities with streamed data and thus the composition and advanced composition theorems apply, according to which privacy disclosure scales with length $T$, respectively, as ${\mathcal O}(T)$ and ${\mathcal O}\left(\sqrt{T}\right)$~\cite{kairouz2015composition}. However, in streaming, subsequent data points are typically neighbors, whereas subsequent queries are typically random points~\cite{ali2005hippocratic}. We can expect that this influences the disclosure of private information, though it is not immediately clear whether it increases or decreases the threat. \changed{In any case, according to the proposal for ``Principles of Hippocratic Stream Data Management Systems''~\cite{ali2005hippocratic}, a privacy-preserving processing of streamed data must ensure}
\begin{enumerate*}
    \item Purpose Specification,
    \item Consent,
    \item Limited Collection,
    \item Limited Use,
    \item Limited Disclosure,
    \item Limited Retention,
    \item Accuracy (in terms of sampling, quantization, and delay),
    \item Safety,
    \item Openness (for the data owner to access information about themself),
    \item Compliance (for the data owner to verify compliance with privacy-preserving principles),
    \item Faithful Representation (of the underlying, accurate data), and
    \item Minimum Quality of Service (QoS).
\end{enumerate*} The current work focuses on principle 7 by quantifying disclosure in streamed systems, as well as the evaluation of the limitations of purposes \numrange{3}{6}.

The statistics of rank-order data, such as that discussed in this work, may be studied using \emph{order statistics} (e.g.~\cite[Section 5.5]{casella2024statistical}). Such tools can be used to determine, for example, the probability that an input is the $n$th largest in a set, or the probability distribution of the same. More specifically, assessing the accuracy of biometric identification has been characterized with the beta-binomial distribution~\cite{schuckers2003using}. However, as far as the authors know, the beta-binomial distribution has not been applied to real-world data, and speech applications in particular, and its application has mainly been from a biometric identification rather than a privacy-preserving perspective~\cite{dass2006validating,mitra2007statistical}. A crucial difference between the two approaches is that biometric identification concerns the probability of correct identification (and its confidence intervals). In contrast, in privacy-preserving processing, we are concerned with the amount of information revealed about a user (e.g., in bits). Notably, the recognition probability depends on the size of the template database, whereas information disclosure should be independent of that.

The theoretical analysis of privacy has become almost synonymous with \emph{differential privacy}~\cite{dwork2014algorithmic}. It is a mathematical framework for ensuring that the analysis of a dataset does not disclose information about individual data points, such as individual users. A central tool for ensuring such protection is adding noise to the outputs (or intermediate points) of the protected system. While the framework gives strong tools for ensuring privacy, it has been adopted in only a few speech processing applications~\cite{feng2022emo,Shamsabadi2022diff,feng22b_interspeech,shoemate2022sottovoce}.

Metrics for evaluating privacy-preserving speech processing technologies include 
\begin{enumerate*}
    \item the \emph{Equal Error Rate (EER)}, which is defined as the threshold where the rates of False Positives and False Negatives are equal~\cite{jahangir2021speaker,maouche20_interspeech},
    \item the Privacy \emph{ZEBRA} metric~\cite{nautsch20_interspeech}, which assesses the average and worst-case protection of a privacy-protection system,
    \label{zebra}
    \item the \emph{application-independent log-likelihood-ratio cost function} $C^{\min}_{llr}$ that generalizes the EER by considering optimal thresholds over all possible prior probabilities and all possible error cost functions~\cite{maouche20_interspeech,brummer2006application},
    \item the \emph{maximum achievable key length from the Receiver Operating Characteristic} (ROC) of an optimal biometric recognition system~\cite{veldhuis2015relation},\label{roc}
    \item the \emph{tandem equal error rate} (t-EER) that jointly evaluates countermeasures and biometric comparators~\cite{kinnunen2023teer},
    \item \emph{linkability} $D^{sys}_{\leftrightarrow}$ computes the overlap in scores between mated (same speaker) and non-mated (different speakers) scores~\cite{gomez2017general},\label{linkability}
    and
    \item \emph{$k$-anonymity}, where a speaker cannot be distinguished within a group of $k$  speakers~\cite{sweeney2002k}.
\end{enumerate*}
Out of these, only Metrics \ref{zebra} and \ref{roc} provide disclosure in terms of entropy (bits) and are thus compatible with the theory of differential privacy. Similarly, only Metric \ref{linkability} explicitly considers the distribution of \emph{other} speakers. None of them combines these two beneficial properties.

\begin{table*}[t]
    \caption{Scenario of use for the privacy-preserving application following~\cite{rahman24_spsc}.}
    \setstretch{1.0}
    \centering\small
    \begin{tabular}{p{1.37cm}p{5.7cm}|p{1.37cm}p{7.56cm}}
    \multicolumn{2}{c|}{\textbf{Attacker Model}} &
    \multicolumn{2}{c}{\textbf{Protector Model}} \\\hline
    Objective & Reidentify user from anonymized sample. & Objective & 
    \textbf{Defense objective:} Protect identity of user.\newline 
    \textbf{Utility objective:} \strut Retain the utility of the downstream application.\\\hline
    Opportunity & Access to anonymized sample (public data).\strut & Opportunity & An anonymization method is available. \\\hline
    \strut Resources & Database of biometric templates (found data). & \strut Resources & None. \\
    \end{tabular}
    \label{tab:scenario}
\end{table*}

\def\db{%
            \draw[line width=.022cm] (0,.25) ellipse (.4cm and .2cm);
            \draw[line width=.022cm] (0,-.25) ellipse (.4cm and .2cm);
            \draw[fill=white,color=white] (-.4,0) -- (.4,0) -- (.4,-.25) -- (-.4,-.25) -- (-.4,0);%
            \draw[line width=.022cm] (-.4,-.25) -- (-.4,.25);%
            \draw[line width=.022cm] (.4,-.25) -- (.4,.25)}%

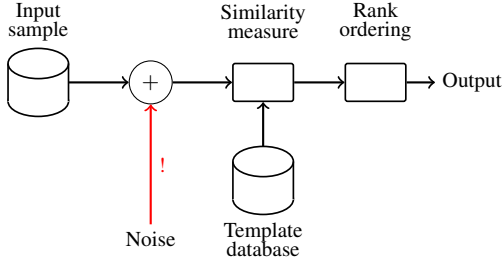
\begin{figure}[t]
    \setstretch{0.8}
    \centering
    \footnotesize
    \begin{tikzpicture}
        \node[matrix,column sep=.1cm] at (0,0) {
            \node {\parbox{1.2cm}{\centering Input sample}}; 
            &
            & \node{\parbox{1.2cm}{\centering Similarity measure}}; 
            & \node{\parbox{1.2cm}{\centering Rank ordering}}; \\
            \db; \node[minimum width=.8cm] (private) {}; 
            & \node[circle, draw] (add) {$+$};
            & \node[draw,rounded corners=.03cm,minimum width=.8cm,minimum height=.5cm,line width=.022cm] (simil) {}; 
            & \node[draw,rounded corners=.03cm,minimum width=.8cm,minimum height=.5cm,line width=.022cm] (rank) {}; 
            & \node (out) {Output};
            \\\node{~};\\ \node{~};\\ 
            && \db; \node[minimum width=.8cm] (db) {}; 
            \\
            & \node (noise) {\parbox{1.2cm}{\centering Noise}}; 
            & \node {\parbox{1.2cm}{\centering Template database}}; 
            \\
        };
        \draw[->,thick] (private) -- (add);
        \draw[->,thick] (add) -- (simil);
        \draw[->,thick] (simil) -- (rank);
        \draw[->,thick] (rank) -- (out);
        \draw[->,thick,red] (noise) -- node[midway,right]{!} (add);
        \draw[->,thick] ($(db)+(0,.45)$) -- (simil);
    \end{tikzpicture}
    \caption{The \emph{threat model} for evaluating the accuracy of biometric recognition, where the input is corrupted by noise, thwarting the accuracy of the similarity measure to templates in the database and thus altering their rank ordering. The unfavorable insertion of noise is highlighted with a red line and an exclamation mark.}
    \label{fig:idobj}
\end{figure}

\section{Threat Models}\label{sec:threats}

The threat model of biometric identification we use is illustrated in \cref{fig:idobj}. We assume that all modules of the identification application are functioning correctly; however, the input sample is noisy, which reduces the accuracy of the similarity measure and corrupts the rank order. This is functionally equivalent to a noise-free input but with an unreliable similarity measure that may not always return a correct answer.

\begin{figure}[t]
    \setstretch{0.8}
    \centering
    \footnotesize
    \begin{tikzpicture}

        \node[matrix,column sep=.1cm] at (0,0) {
            \node {\parbox{1.2cm}{\centering Private data}}; 
            & \node{Anonymization}; 
            & \node{\parbox{1.2cm}{\centering Public data}}; 
            & \node{\parbox{1.5cm}{\centering Downstream application}}; \\
            \db; \node[minimum width=.8cm] (private) {}; 
            & \node[draw,rounded corners=.03cm,minimum width=.8cm,minimum height=.5cm,line width=.022cm] (anon) {}; 
            & \db; \node[minimum width=.8cm,minimum height=.9cm] (public) {}; 
            & \node[draw,rounded corners=.03cm,minimum width=.8cm,minimum height=.5cm,line width=.022cm] (asr) {}; 
            \\
            \node{~};\\ \node{~};\\ 
            & \db; \node[minimum width=.8cm] (found) {};
            & \node[draw,red,rounded corners=.03cm,minimum width=.8cm,minimum height=.5cm,line width=.022cm] (attack) {!}; 
            & \node[red] (out) {\parbox{1.4cm}{\centering Private\newline information}};
            \\
            & \node{\parbox{1.6cm}{\centering Found data}}; 
            & \node[red]{\parbox{1.2cm}{\centering Attacker}};\\
        };        

        \draw[->,thick] (private) -- (anon);
        \draw[->,dashed,thick] (anon) -- (public);
        \draw[->,thick,dashed] (public) -- (asr);
        \draw[->,dashed,red,thick] (public) -- (attack);
        \draw[->,red,thick] (found) -- (attack);
        \draw[->,red,thick] (attack) -- (out);
        
    \end{tikzpicture}
    \caption{The \emph{threat model} for evaluating privacy-preserving anonymization  following~\cite{maouche20_interspeech}, where private data is anonymized to remove private information, and the anonymized data is shared publicly. An attacker uses any available (found) data and anonymized public data to infer private information contrary to the users' preferences. The anonymized data flow is indicated by dashed lines, and the attack by red lines and an exclamation mark.}
    \label{fig:anonobj}
\end{figure}
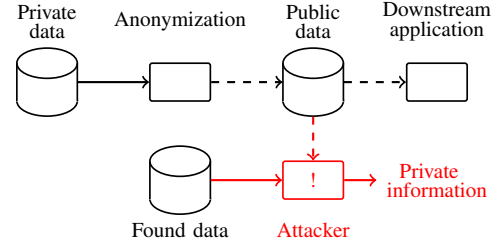

The threat model of the privacy-preserving applications is illustrated in \cref{fig:anonobj}, based on~\cite{maouche20_interspeech}. The attacker and protector's objectives, opportunities, and resources are further characterized in a scenario of use in \cref{tab:scenario}, following~\cite{rahman24_spsc}. We assume that private data, including biometric information, is anonymized before it becomes public. Thus, the attacker has reduced opportunities to re-identify the input sample. However, the attacker can gain aid from access to ``found data'', corresponding to any information source. Observe that from the attacker's perspective, \cref{fig:anonobj} is identical to \cref{fig:idobj}, as ``found data'' and ``anonymization'' in \cref{fig:anonobj} correspond to the ``template database'' and ``noise'' in \cref{fig:idobj}. The attacker's objective is the opposite of the user's. The two threat models are thus, for our purposes, otherwise identical except that the optimization objectives have opposite signs. For the quantification of private information, this sign does not make a difference, and we can use \cref{fig:idobj} as our threat model without reducing generality.

\section{Similarity Rank Disclosure}\label{sec:methods}
Suppose we have a dataset of speaker templates, an input speech signal, and a similarity metric. Assuming one of the templates corresponds to the input speech signal, we can calculate the similarity between all templates and the input. If the data and similarity metrics are accurate, then the most similar template should correspond to the true speaker. However, when the similarity metric or data is imperfect, and especially when speakers have similar voices, the most similar template may not match the true speaker. We can thus rank order the templates according to their similarity metrics. Similar voices will then have a low rank and dissimilar voices a high rank. 

The central quantity of interest is then the location $k\in[1,N]$ of the true speaker in the similarity rank in a set of $N$ templates. The probability that the true speaker will have similarity rank $k$ is then $p(k)$. \Cref{fig:samplehist} illustrates possible similarity rank distributions. Pane (a) corresponds to a typical situation where the lowest (most similar) ranks are most likely. Noisy data and a poor similarity measure will give a wider distribution (pane (b)). A good similarity metric and data will give a narrower distribution centered at the low end (pane (c)). The shape of the distribution also bears information about the setup. At one extreme case, if the non-matching ranks ($\textrm{rank} > 1$) are equally likely (pane (d)), it means that a non-matching observations bear a minimum of information about the true speaker. The other extreme is when $k$ lowest ranks have equal probability (pane (e)), giving $k$-anonymity for the speaker~\cite{sweeney2002k}. Theast case, pane (f), illustrates a similarity rank histogram where higher ranks have a higher probability. This anomalous case indicates that the similarity rank is functioning incorrectly, since dissimilarity is a better indication of identity than similarity.

\subsection{Modelling Rank Distribution}
Suppose we have a database of user templates, $c_k\in{\mathbb R}^{M\times 1}$, $k\in[1,N]$, and an input $x\in{\mathbb R}^{M\times 1}$ where $M$ is the dimensionality of the space. We can calculate the distance between the input and each of the templates as $d_k = d(c_k,x)$. Suppose $c_l$, for some $l\in[1,N]$ matches the true identity of the input. Let the probability that the distance to the true identity $d_l$ is smaller than the distance to some other template $d_k$ be 
\begin{equation}
    P[d_l < d_k] = \theta_{k}.
\end{equation}
The probability that $d_l$ is the smallest of all $N$ templates is then 
\begin{equation}
    P\left[\changed{\cap_{k\neq l} d_l < d_k}\right] = \prod_{k\neq l} \theta_k.
\end{equation}
With a large database size $N$, this becomes increasingly improbable. We can thus instead consider the probability that
$c_l$ is the $j$th best match.

\changed{Suppose} the $\theta_k$'s are drawn from a beta-distribution with parameters $\alpha,\beta$. \changed{This choice gives reasonable flexibility to accommodate a wide variety of datasets with only two parameters.}
The probability that $c_l$ is the $j$th best match follows the beta-binomial distribution~\cite{casella2024statistical}
\begin{equation}\label{eq:betabino}
    \gamma_j:=P\left[j = \changed{\mathrm{rank}\left(d_l, d_k\right)}\right] = \binom{N}{j}\frac{B(j+\alpha,N-j+\alpha)}{B(\alpha,\beta)},
\end{equation}
where \changed{$\mathrm{rank}\left(d_l, d_k\right)$ gives the rank of $d_l$ in the set $d_k$ and} $B(\cdot)$ is the beta function
\begin{equation}
    B(\alpha,\beta) = \frac{\Gamma(\alpha)\Gamma(\beta)}{\Gamma(\alpha+\beta)}.
\end{equation}
From the perspective of an attacker for whom the true identity is not known, \cref{eq:betabino} gives the posterior probability that the true identity is $j$. Assuming all identities were equally likely before the observation, the prior probability is $\frac1N$ \changed{and correspondingly, encoding this information requires $\log_2N$ bits. Similarly, encoding the information $\gamma_j$ that the true speaker is the $j$th most similar requires $-\log_2\gamma_j$ bits.}
The disclosure of private information or the \emph{rank order disclosure} is thus \changed{the difference}
\begin{equation}
    \epsilon_j := \changed{-\log_2\gamma_j-\log_2N.}
\end{equation}

Note that this disclosure provides information about all identities in the database, regardless of whether they match the input or not. However, the disclosure may or may not support identification. A positive disclosure, $\epsilon_j>0$, means that the posterior probability that the input corresponds to speaker $j$ is larger than before the observation. Correspondingly, a negative disclosure, $\epsilon_j<0$, means that the probability that the observation is speaker $j$ has diminished.

\subsection{Parameter Estimation}\label{sec:paramest}
To use \cref{eq:betabino}, we need to estimate the parameters $\alpha,\beta$ from data. Our informal experiments have shown that the trivial method of moments approach generally yields nonsensical values (negative $\alpha,\beta$)~\cite{wilcox1979estimating}. With a histogram of observations, $h_k$, $k\in[1,N]$, normalized to the empirical probability distribution ${\tilde p_k}:=\frac{h_k}{\sum_k h_k}$, we therefore optimize $\gamma_k$ using the objective functions:\\[1pt]

\noindent
\begin{tabular}{ll@{}}
    log-likelihood (LL) & $\min\limits_{\alpha,\beta}-\sum\limits_k {\tilde p_k} \log \gamma_k$ \\ [1em]
    mean square (MS) & $\min\limits_{\alpha,\beta} \sum\limits_k \left({\tilde p_k} - \gamma_k\right)^2$\\ [1em]
    \parbox{3cm}{weighted mean\newline square (WMS)} & $\min\limits_{\alpha,\beta} \sum\limits_k {\tilde p_k}\left({\tilde p_k} - \gamma_k\right)^2$\\ [1em]
    \parbox{3cm}{rank-weighted mean\newline square (RWMS)} & $\min\limits_{\alpha,\beta} \sum\limits_k e^{-k}\left({\tilde p_k} - \gamma_k\right)^2$\\ [1em]
    \parbox{3cm}{constrained log-\newline likelihood (CLL)} & $\min\limits_{\alpha,\beta}-\sum\limits_k {\tilde p_k} \log \gamma_k + 10^5 \left({\tilde p_1}-\gamma_1\right)^2$. \\ [1em]
\end{tabular}\\[1pt]

The goodness-of-fit for the estimated models are evaluated with two approaches; the base-2 Kullback-Leibler (KL) divergence and $\log_2$-match for the first rank:\\[1pt]

\noindent
\begin{tabular}{ll}
     KL-divergence &  $\sum_k {\tilde p_k} \log_2 \frac{{\tilde p_k}}{\gamma_k}$ \\ [0.5em]
     rank-1 match & $\left|\log_2 \frac{{\tilde p_1}}{\gamma_1}\right|$
\end{tabular}\\[1pt]

\noindent
where ${\tilde p_k}$ and $\gamma_k$ are the normalized histogram (empirical probability distribution) and beta-binomial model, respectively. The KL-divergence thus quantifies overall performance, while the rank-1 match quantifies goodness-of-fit for the most important rank, i.e., ${\mathrm{rank}}=1$.

\subsection{Rank Disclosure Statistics}\label{sec:stats}
Beyond the disclosure of individual observations, it is essential to characterize the statistical properties of disclosure. 
In particular, both the mean and worst-case disclosure are of interest~\cite{vali2024porcupine, nautsch20_interspeech}. With $k$ as the rank of the true speaker among $N$ speaker and with the empirical probability distribution ${\tilde p_k}$ (or its corresponding model $\gamma_k$) the pertinent statistics are:\\[1pt]

\noindent
\begin{tabular}{ll@{}}
   individual disclosure & $\epsilon_k := \changed{-}\log_2N-\log_2{\tilde p_k}$\\ [0.5em]
   mean disclosure (MeanD) & $\mu:= \sum_k {\tilde p_k}\epsilon_k$ \\ [0.5em]
   identification rate (IdR) & ${\tilde p_1}$ \\ [0.5em]
   \parbox{4cm}{\strut standard deviation of mean disclosure (StDD)\strut} & $\sigma := \sqrt{ \sum_k {\tilde p_k}\left(\epsilon_k-\mu\right)^2 }$ \\ [1em]
   maximum disclosure (MaxD) & $\max_k \epsilon_k$ \\ [0.5em]
   rank spread (Spread) & $\frac1N{\mathrm{count}_{k\in[1,N]}}\left({\tilde p_k} > \frac1N\right).$
   \\ [0.5em]
\end{tabular}\\[1pt]

Here, the \emph{rank spread} encompasses the proportion of bins where the (empirical) probability is above the equal-probability level (i.e., pure chance), using the $\mathrm{count}_k()$ function that counts the number of true expressions.

\Cref{fig:samplehist} illustrates the mean disclosure (MeanD) and rank spread of sample histograms.

\subsection{Summary}
The proposed metric quantifies the disclosure of personally identifying information (PII) from similarity rank information. For a similarity measure $S(x,y)$, the required steps are
\begin{enumerate}
    \item For every test input $x_k$
    \begin{enumerate}
        \item Evaluate the similarity $S(x_k,y_h)$ to every database template $y_h$.
        \item Rank order similarity scores such that the most similar is first.
        \item Store the rank of the true match in the histogram.
    \end{enumerate}
    \item Normalize histogram to obtain the empirical probability distribution.
    \item If the distribution is sufficiently dense (plenty of data), continue to step \ref{stats}.
    \item Fit a beta-binomial distribution to the empirical probability distribution (see \cref{sec:paramest}).
    \item Evaluate the disclosure statistics for the histogram or beta-binomial model (see \cref{sec:stats}).\label{stats}
\end{enumerate}

\section{Data}\label{sec:data}
\newcommand{\Librivox}{LibriLong}

We use three speech corpora for our experiments: the Fisher English Training Speech Part 2 corpus~\cite{cieri2005fisherpart2}, VoxCeleb~\cite{nagrani17_interspeech}, and a new purpose-built dataset, \Librivox, based on the LibriVox free public domain audiobooks\footnote{\url{https://librivox.org/}}, similar to LibriSpeech~\cite{panayotov2015librispeech}. The Fisher dataset contains \qty{975}{\hour} of spontaneous, conversational speech at a sampling rate of \qty{8}{\kilo\hertz}. There are \num{5848} conversations of \qty{10}{\minute} each, with the two speakers in separate channels, which gives a total of \num{11696} unique speakers. VoxCeleb contains \num{1251} speakers, each with an average of \num{116} utterances, each of length \qty{8.2}{\second} on average\changed{~\cite{nagrani17_interspeech}}. Speaker identities in Fisher and VoxCeleb thus correspond to \qty{13.5}{\bit} and \qty{10.3}{\bit} of information, respectively.

The \Librivox{} dataset contains 64 excerpts of audiobooks in English, each of at least \qty{2}{\hour} in length, sampled at \qty{16}{\kilo\hertz}. Speaker identities in \Librivox{} thus contain \qty{6}{\bit} of information. The purpose of the dataset is to allow evaluation of privacy threats in long recordings.
Excerpts were randomly chosen from the ``general fiction'' category of LibriVox, such that 
a single person read the whole excerpt, and the excerpt is of at least \qty{2}{\hour} in length.
The distribution of perceived genders of speakers (as judged by the current authors) is 31 male and 33 female.
One candidate was replaced because the fundamental frequency distribution was informally found to differ significantly between chapters of the audiobook. 
This dataset is published openly\footnote{\url{https://dx.doi.org/10.21227/6gvw-vn30}}.
To allow comparison between the Fisher and \Librivox{} datasets, in our experiments, \Librivox{} was resampled to \SI{8}{\kilo\hertz}. \Librivox{} is read text, and the linguistic content thus relates to the author of the audiobook rather than the speaker. Consequently, we did not use \Librivox{} for linguistic analysis.

\section{Speaker features}\label{sec:features}
We characterize speakers with four different features,
\begin{enumerate*}
    \item the fundamental frequency distribution,
    \item a phone embedding distribution, 
    \item a speaker identity embedding, and
    \item a text style embedding.
\end{enumerate*}
This provides a range of abstraction levels for the personally identifying information (PII) of speakers. Notably, the aim is not to develop new or improved features, but we chose standard approaches to demonstrate the use of the proposed disclosure statistics.

\begin{figure*}[t!]
    \centering
    \includegraphics[width=\textwidth]{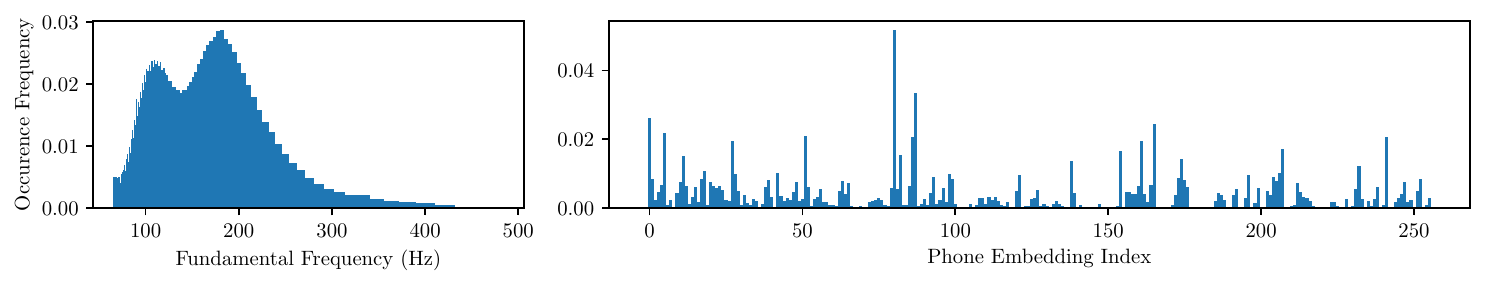}\vspace{-.4cm}
    \caption{Histograms of fundamental frequencies and phone embeddings of all speakers in the Fisher dataset.}
    \label{fig:joint_histogram}
\end{figure*}

The fundamental frequency of a speech sample is analyzed with the YIN algorithm~\cite{de2002yin} using the Torch-YIN implementation\footnote{\url{https://github.com/brentspell/torch-yin}}. We limit the frequency range of the fundamental to \qtyrange[range-units = single]{65}{450}{\hertz}. This implementation estimates pitch lags as integers, which gives \num{107} distinct fundamental frequencies. The stride is \qty{10}{\milli\second} such that each \qty{30}{\second} segment contains up to approximately \num{3000} measurements, but since non-voiced speech frames are discarded, the actual number of measurements per segment is lower. The histogram of fundamental frequencies over the entire Fisher dataset is illustrated in \cref{fig:joint_histogram}. The histograms $h_k$ are normalized to the corresponding empirical probability distribution, $\tilde p_k=\frac{h_k}{\sum_k h_k}$.

Phone embeddings are extracted from a VQ-VAE-based voice conversion algorithm~\cite{Niekerk2020,vali2023interpretable}. We trained the encoder, decoder, and space-filling vector quantizer of~\cite{vali2023interpretable} on the \textit{ZeroSpeech 2019 Challenge} English dataset~\cite{zerospeech_dataset} with the same hyperparameters as in~\cite{vali2023interpretable}, while the speech files were resampled to \SI{8}{\kilo\hertz} to train the model on a similar sampling rate as the Fisher dataset. The quantization of the phone embeddings is done over 256 codebook vectors (i.e., 256 bins). Each \qty{10}{\minute} conversation of Fisher is split into twenty \qty{30}{\second} chunks. For each chunk and each speaker, the silence and the frames shorter than \qty{0.1}{\second} speech are skipped for phone extraction. The histogram of phone embeddings over the entire Fisher dataset is illustrated in \cref{fig:joint_histogram}. Again, the histograms $h_k$ are normalized to the corresponding empirical probability distribution, $\tilde p_k=\frac{h_k}{\sum_k h_k}$.

Speaker identity embeddings are extracted from each audio file in VoxCeleb using the standard baseline for speaker recognition, ECAPA-TDNN~\cite{desplanques20_interspeech}, using its pre-trained SpeechBrain implementation~\cite{speechbrain}. To get an equal number of observations for each speaker, we use the \num{45} first files of each speaker as the observations. The average EER for this dataset was \qty{0.76}{\percent}, which is low enough to confirm that ECAPA-TDNN works properly for this dataset.

The linguistic features are quantified with Linguistic Inquiry and Word Count (LIWC). This computerized text analysis method is widely used to analyze emotional, cognitive, and structural components in written and spoken language~\cite{tausczik2010psychological,boyd2022development}. We use all linguistic features provided by the LIWC-22 dictionary, except those specifically measuring the text content domain, as the conversation topics were not freely chosen in the Fisher dataset. Since our data consists of transcripts, we also consider conversational features such as disfluencies and the frequency of filler words. Combining these elements, we construct an \num{80}-dimensional feature vector. We split the text of the Fisher dataset of each speaker into \num{5} segments of \qty{2}{\minute}. Utterances were assigned to segments based on the location of segment centers: $T_{center}=\frac12\left(T_{start}+T_{end}\right)$. 

\section{Experiments}\label{sec:experiments}
The purpose of our experiments is to
\begin{enumerate*}
    \item demonstrate the usefulness of the proposed disclosure measures by applying them to different categories of personally identifiable information (PII), 
    \item get a first impression of the magnitude of PII disclosure in those categories, 
    \item evaluate the applicability of the proposed beta-binomial model to both dense and sparse histograms,
    and
    \item study disclosure in streamed applications. 
\end{enumerate*}

We construct experiments as follows: Let the number of speakers be $N$, the number of observations per speaker be $K$, and each observation has $M$ dimensions. Let $T,D$ be the number of observations of input samples and the database templates, respectively, and the starting point for input samples be $S$. The indices for samples are then defined as
\begin{equation}
    \begin{matrix}
        \text{input samples} & S +[0\dots T) \mod K\\
        \text{database templates} & S+T+\Delta+ [0\dots D) \mod K ,
    \end{matrix}
\end{equation}
where $\Delta=\frac12(K-S-T)$. This gives non-overlapping segments for the input and database with maximum separation. By iterating $S$ over $[0,K)$, we obtain data similar to an $S$-fold cross-validation experiment.

\begin{figure*}[t!]
    \includegraphics[width=\textwidth]{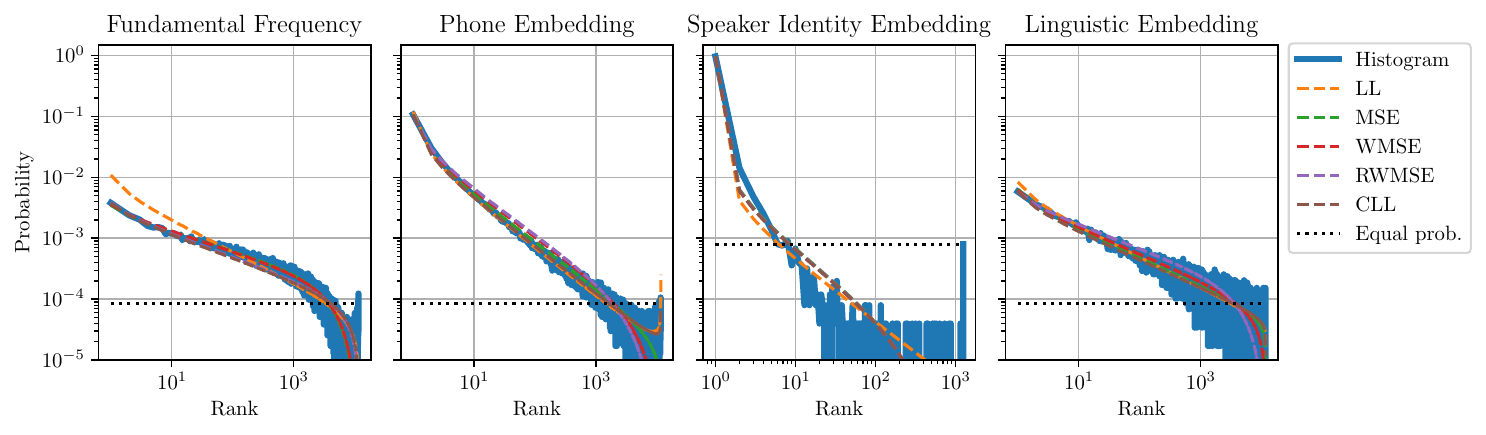}\vspace{-.4cm}
    \caption{The empirical probability distribution of similarity rank. The dashed lines indicate beta-binomial models fitted to the data, and the dotted black line indicates the level of equal probability.}
    \label{fig:jointhist}
\end{figure*}

\Cref{fig:jointhist} illustrates the similarity rank histograms of the fundamental frequency, phone, speaker, and linguistic embeddings with $T=1,\,D=1$. The histograms are plotted in a log-log graph to focus attention on the lowest ranks and to visualize probabilities of different orders of magnitude. We observe that all four plots have a similar shape; the histograms are more or less linear in the log-log domain, except for the lowest ranks, which bend upward.
We can furthermore see that the speaker identity embedding has the highest probability for a match ($\textrm{rank}=1$), followed by phone and linguistic embeddings, as well as the fundamental frequency. Observe that this comparison should be taken as indicative only, since the speaker identity ranks are calculated from a corpus with a smaller number of speakers, and the numbers are thus not directly comparable.

The disclosure statistics for each case are presented in \cref{tab:disclosurestats}. The mean disclosure and identification rates confirm the observation from~\cref{fig:jointhist} that speaker identity embeddings contain the most accurate information. Interestingly, the maximum disclosure is higher for the phone than the speaker embedding. However, since VoxCeleb contains only \qty{10.3}{\bit} of speaker identity information, that will limit the disclosure. We can thus expect that the maximum disclosure from the speaker identity embedding is higher than that. The spread for the disclosure from the fundamental frequency and linguistic embeddings is approximately \qty{28}{\percent}. This indicates the proportion of speakers whose recognition probability will, on average, increase for every observation. Observe that these are predominantly false positives, where the posterior probability increases for speakers similar to the test subject. The spread for phone embeddings is significantly smaller, and even smaller for the speaker identity embedding. 

The Equal Error Rates (EER) of each feature are largely inversely proportional to the mean and maximum disclosure as well as identification rate. The only apparent deviation is the phone embedding, where the EER is higher than for the fundamental frequency; yet, the disclosures and identification rate are also higher. 
\medskip

\begin{table}[t!]
    \caption{Disclosure statistics of similarity rank histograms, beta-binomial model optimized with CLL and the baseline metric equal error rate (EER) corresponding to the fundamental frequency (F0), phone, speaker identity, and linguistic embeddings.}
    \centering
    \sisetup{
        table-alignment-mode= format,
        table-number-alignment = center,
        tight-spacing=true
        }
    \footnotesize
\begin{tabular}{l|SSSS}
\multicolumn{5}{l}{Histograms}\\
& {F0} & {Phone} & {Identity} & {Linguistic} \\\hline
MeanD (bits) & 0.99  & 2.45  & 9.97  & 0.66  \\
IdR (\%) & 0.38  & 10.77  & 96.74  & 0.58  \\
MaxD (bits) & 5.48  & 10.30  & 10.24  & 6.10  \\
StDD (bits) & 1.57  & 3.87  & 1.57  & 1.44  \\
Spread (\%) & 28.15  & 13.47  & 0.48  & 28.22  \\
\hline\multicolumn{5}{l}{}\\
\multicolumn{5}{l}{Model optimized with CLL}\\
& {F0} & {Phone} & {Identity} & {Linguistic} \\\hline
MeanD (bits) & 0.59  & 2.27  & 9.89  & 0.41  \\
IdR (\%) & 0.38  & 10.67  & 96.75  & 0.57  \\
MaxD (bits) & 5.47  & 10.29  & 10.24  & 6.07  \\
StDD (bits) & 1.31  & 3.79  & 1.99  & 1.29  \\
Spread (\%) & 34.95  & 14.00  & 0.64  & 28.77  \\
\hline\multicolumn{5}{l}{}\\
\multicolumn{5}{l}{Baseline metric}\\\hline
EER (\%) & 32.30 & 37.78 & 0.76 & 39.10 \\
\hline
\end{tabular}
    \label{tab:disclosurestats}
\end{table}

To each of the empirical probability distributions (i.e., normalized histograms), we fitted the beta-binomial distribution using the loss functions listed in~\cref{sec:paramest}. The models are illustrated in~\cref{fig:jointhist}. We observe that though the log-likelihood is the standard approach to fitting distributions, in most cases, it provides a poor fit for the lowest ranks. In the current application, the lowest ranks are the most important, as there the disclosure is the highest. This warranted informal experimenting with different losses (see \cref{sec:paramest}), the most promising of which are included in this paper.

The goodness-of-fit measures of each feature and loss are listed in~\cref{tab:gof}. We observe that the log-likelihood (LL) obtains the best (smallest) KL-divergence in all cases except one, where it is very close. This is natural since the log-likelihood and KL-divergence differ only by a constant scalar offset. The constrained log-likelihood (CLL), which penalizes for a mismatch in the rank-1 value, also obtains the best (smallest) rank-1 match, except in two cases where it is very close. As the constrained log-likelihood (CLL) is reasonably good also concerning the KL-divergence, we conclude that it is the best loss for our purposes.\smallskip

The disclosure statistics for the beta-binomial model in the four cases are listed in \cref{tab:disclosurestats}.
Overall, the statistics roughly align with those extracted from the raw histograms. The identification rates (IdR) and maximum disclosure (MaxD) all match within \qty{1}{\percent} difference. The mean disclosures (MeanD) have at most a \qty{0.4}{\bit} difference. Observe that such a difference can be expected as the model is an approximation. However, we cannot unequivocally say that either the raw data or the model would be better or more accurate than the other, since data is noisy and the model is a simplification. Still, the differences in disclosures indicate the accuracy that can be expected of these metrics. In other words, with the current measurement setup, the mean disclosure metric can have an error of the order of \qty{0.5}{\bit}.
\medskip

\begin{table}[t!]
    \centering
    \caption{Goodness-of-fit measures for the beta-binomial model optimized with different loss functions (see \cref{sec:paramest}) and for each of the speech features. Smaller is better. The best loss for each combination of feature and metric is in bold. }
    \label{tab:gof}
    \sisetup{
        table-alignment-mode= format,
        table-number-alignment = center,
        tight-spacing=true
        }
    \resizebox{\columnwidth}{!}{
\begin{tabular}{l|l|S[table-format=2.2e1]S[table-format=2.2e1]S[table-format=2.2e1]S[table-format=2.2e1]}
Metric & Loss & {F0} & {Phone} & {Identity} & {Linguistic} \\\hline
\multirow{5}{*}{\rotatebox{90}{\parbox{1.5cm}{\centering KL-div. (bits)}} }
 & LL & \bfseries 1.66e-01 & 4.97e-02 & \bfseries 3.94e-02 & \bfseries 1.57e-01 \\
 & MSE & 3.94e-01 & 2.12e-01 & 6.70e-02 & 1.66e-01 \\
 & WMSE & 3.50e-01 & 7.97e-01 & 6.53e-02 & 2.64e-01 \\
 & RWMSE & 1.97e-01 & 1.25e+00 & 6.65e-02 & 6.78e-01 \\
 & CLL & 1.96e-01 & \bfseries 4.87e-02 & 6.54e-02 & 1.70e-01 \\
\hline
\multirow{5}{*}{\rotatebox{90}{\parbox{2cm}{\centering Rank-1 match (bits)}} }
 & LL & 1.52e+00 & 1.45e-01 & 1.11e-03 & 5.28e-01 \\
 & MSE & 1.33e-01 & 1.32e-02 & 1.40e-03 & 1.37e-01 \\
 & WMSE & 7.72e-02 & 4.50e-02 & \bfseries 1.92e-04 & 6.92e-02 \\
 & RWMSE & 2.98e-02 & 1.25e-02 & 3.46e-04 & \bfseries 1.66e-02 \\
 & CLL & \bfseries 1.63e-03 & \bfseries 7.94e-03 & 1.03e-03 & 2.65e-02 \\
\hline
\end{tabular}
}
\end{table}

\begin{figure}[t!]
    \centering
    \includegraphics[width=\columnwidth]{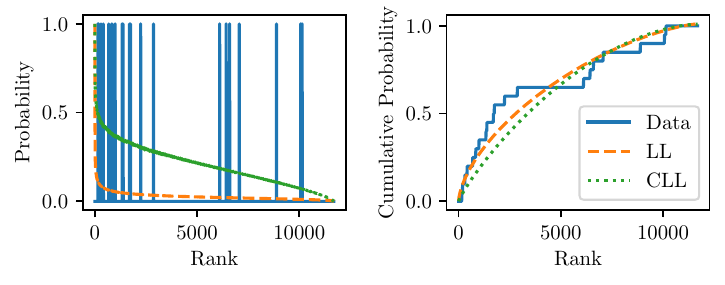}\vspace{-.4cm}
    \caption{The histogram/probability distribution (left pane) and the empirical cumulative probability distribution (right pane) for the similarity rank obtained from the fundamental frequency of one speaker in the Fisher dataset, as well as the corresponding beta-binomial models fitted with two different losses, LL and CLL. The probability distributions of the models are scaled to unity for improved visibility. }
    \label{fig:individualhist}
\end{figure}

The above histograms were accumulated over all speakers. Individual speakers may, however, have different characteristics, making them easier or harder to identify. We may thus expect that the similarity rank histograms are significantly different for individual speakers. \Cref{fig:individualhist} illustrates the histogram and the corresponding empirical cumulative distribution (scaled to unity) of an example speaker. We use linear axes since the sparse histogram is mostly zero, which cannot be displayed in a log-log plot. The histogram is also always only one or zero, such that the empirical cumulative probability distribution serves as a more informative visualization. We included only models fitted with LL and CLL here, as the other models did not always converge.

We observe that the model optimized with the log-likelihood (LL) criteria visually follows the cumulative distribution more closely than the constrained alternative (CLL). A practical issue is that, in many cases, such as in this plot, the histogram value at rank 1 is zero, although we can reasonably expect it to be positive. This means that optimizing CLL will try to force the model to zero at rank-1, even though that contradicts our expectations of the true probability distribution. This introduces a random component to models optimized with CLL. This leaves the log-likelihood loss as the only remaining viable option.
\smallskip

\begin{figure*}[t!]
    \centering
    \includegraphics[width=\textwidth]{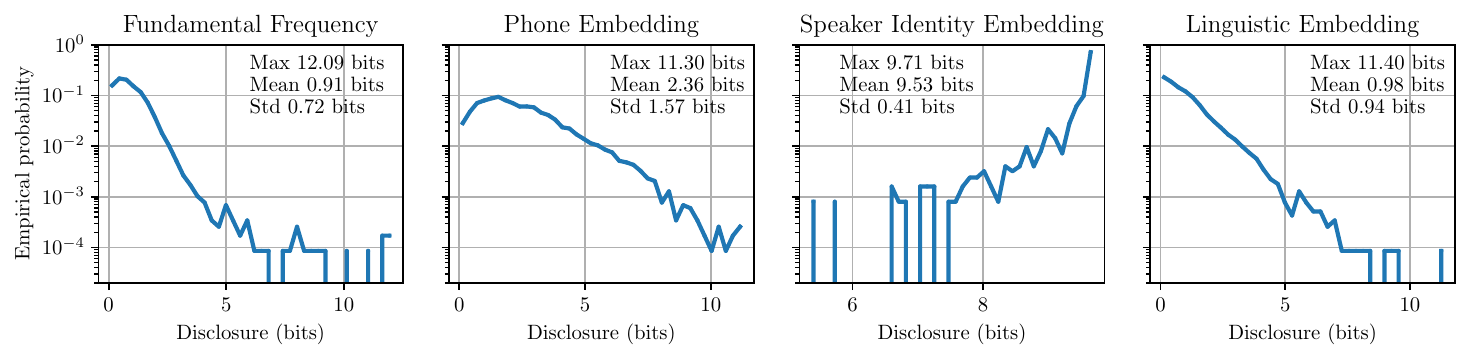}\vspace{-.4cm}
    \caption{The empirical probability distribution of the mean disclosure for each feature, calculated from the beta-binomial models fitted with the log-likelihood criteria to the similarity rank histograms. The maximum, mean, and standard deviation (std) of the mean disclosures are included for each feature.  }
    \label{fig:meandisclosure}
\end{figure*}

\Cref{fig:meandisclosure} illustrates the empirical probability distribution of mean disclosure of the beta-binomial models fitted to the individual speakers' histograms. This can be interpreted as the probability distribution of the privacy leak associated with each feature. The maximum disclosure is close to the worst-case for all features, matching the maximum disclosure of the corresponding datasets, which indicates that some speakers are outliers and can be perfectly identifiable from any of the features. This is a concern, as the average disclosures are, in all cases except the speaker identity embedding, reasonably low, in the range \qtyrange[]{0.9}{2.4}{\bit}. In other words, a low mean disclosure does not automatically indicate that all individuals would be protected from identification.
We can, however, see that the distributions of the fundamental frequency, the phone, and linguistic embeddings have a distinct peak around which a majority of individuals are focused. The standard deviations are also reasonably small, below \qty{2}{\bit}. This indicates that a majority of speakers have a similar mean disclosure, and the worst-case scenario concerns only a minority of speakers. The speaker identity embedding peaks at the edge of the range of possible disclosures, indicating that a large proportion of speakers can be perfectly identified. We hypothesize that if the set of speakers and, correspondingly, the range of possible mean disclosures were larger, then the speaker identity embedding would also have a distinct peak. Then all four cases would show a similar shape of the distribution.
\medskip

\begin{figure}[t!]
    \centering
    \includegraphics[width=\columnwidth]{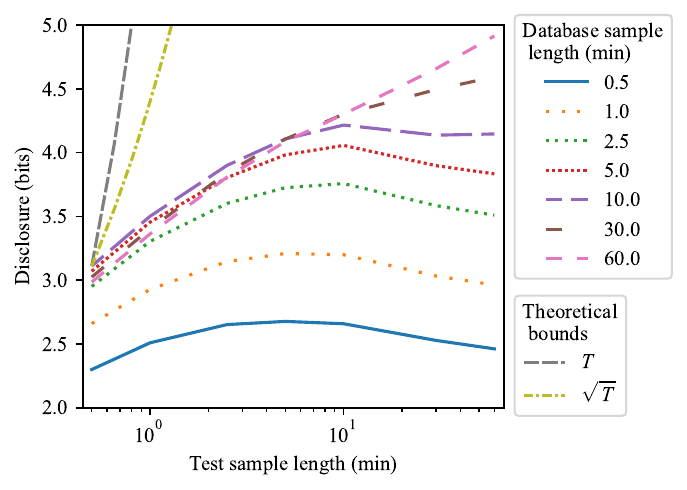}\vspace{-.4cm}
    \caption{Disclosure of the fundamental frequency as a function of the length of segments used for database templates and test samples. The dashed lines show the theoretical limits to scaling of disclosure based on the composition ($T$) and advanced composition ($\sqrt T$) theorems~\cite{kairouz2015composition}.}
    \label{fig:lengthscaling}
\end{figure}

Finally, \cref{fig:lengthscaling} illustrates how disclosure of the fundamental frequency scales with the length of the database samples and test samples. We observe that disclosure initially increases with the length of the database samples and test samples, but counterintuitively, it diminishes slightly at higher lengths. Only database samples of lengths \num{30} and \qty{60}{\minute} depart from this pattern, but we assume that with longer test samples, the pattern would repeat. We hypothesize that the increase in accuracy with longer observations causes the initial increase in disclosure. However, speech features are not stationary due to intra-speaker variance, which causes a bias between the database template and test samples that becomes more pronounced in longer observations. This bias is the likely cause for the reduction in disclosure at higher test sample lengths. Due to symmetry between the test and database samples, the same observations hold with their roles reversed.

According to the basic and advanced composition theorems, disclosure scales with length $T$, respectively, as ${\mathcal O}(T)$ and ${\mathcal O}\left(\sqrt{T}\right)$~\cite{kairouz2015composition},
illustrated by the long dashed and dashed-dotted lines. We observe that the measured disclosure scales orders of magnitude slower than the theoretical bounds.

\section{Discussion}
Quantification is central to evaluating privacy threats. We propose a new metric, named \emph{similarity rank disclosure}, that quantifies the amount of personally identifying information (PII) contained in an observation. The metric measures entropy (bits) and thus aligns with the theory of differential privacy. 

A central outcome of the proposed metric is that the posterior probability of \emph{every} speaker will depend on the observation. In other words, every observation carries information about every speaker. The posterior probability of speakers similar to the test sample will increase, while that of dissimilar speakers will decrease. This insight carries interesting consequences for privacy. If speakers $A$ and $B$ are similar, and $A$ is found to have property $X$, then that increases the probability that speaker $B$ has the same property. In other words, \emph{an observation of speaker $A$ can violate the privacy of speaker $B$, even when speaker $B$ has no interaction with the attacker}. This highlights that privacy is not only about individual users, but it has to be evaluated also on community- and societal levels (cf.~\cite{backstrom2025beyonduser}).

A second outcome is that the posterior of a speaker can decrease due to an observation, even for the true speaker. A decrease in the posterior means that the disclosure has a negative sign and is thus ``negative information''. We can thus categorize observations and their corresponding disclosures as:

\medskip

\noindent
{
    \centering
    \begin{tabular}{l|c|c}
        & \multicolumn{2}{c}{Disclosure} \\\cline{2-3}
        Speaker & Positive & Negative \\\hline
        Match &  True positive & False \changed{negative}\\
        Other &  False positive & True negative
    \end{tabular}\\[.2cm]
    
 }

A third outcome is that while privacy disclosure in streaming applications is, in theory, increasing without bounds with time, the situation is not quite that bleak in practice. First, when the observation is arbitrarily long, the limiting factor for disclosure is the size of the database. Second, intra-speaker and recording environment variance will significantly reduce the potential disclosure. This does not mean that streaming would be ``safe'', but only that the privacy exposure is not as extreme as the theoretical limits would suggest.

\medskip

 In our experiments, we evaluated the disclosure of PII from the fundamental frequency and phone distributions, as well as speaker recognition and linguistic embeddings. These features were selected to represent different levels of abstraction and to demonstrate the properties of the proposed metric. Notably, the analysis of these features was implemented using standard tools without any effort to improve them, to maintain focus on the metric. In particular, it is clear that a great deal more information could be extracted from, for example, the fundamental frequency by analyzing its trajectories over time. Such analysis is left for future studies.

In conclusion, we have demonstrated that the proposed metric can be applied to different speech features. The magnitude of disclosure in each case aligns with the authors' intuitive expectations, in that all features do leak PII, and the speaker recognition embedding, specifically developed to reveal the identity of the speaker, has the highest disclosure. For cases with sparse data, we furthermore proposed a model for the probability distribution. The proposed method is thus applicable to a wide range of speech, text, and other biometric applications that require the evaluation of disclosure of personally identifying information (PII).



\printbibliography[title={References}]

\begin{IEEEbiography}[{\includegraphics[width=1in,height=1.25in,clip,keepaspectratio]{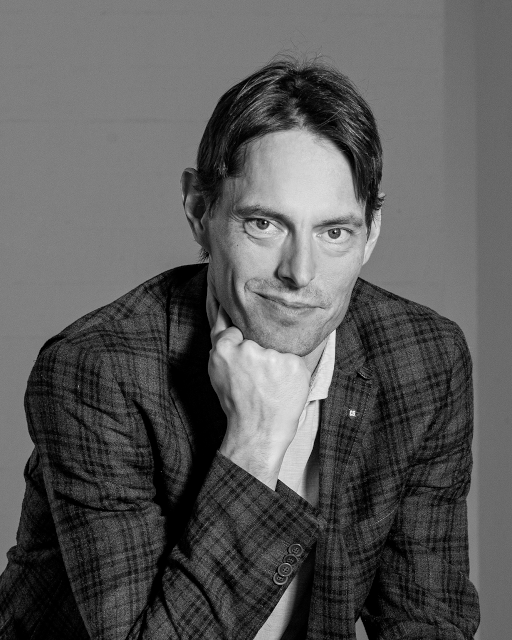}}]{Tom Bäckström} (Senior Member, IEEE) received the master’s and Ph.D. degrees from Aalto University, in 2001 and 2004, respectively, which was then known as the Helsinki University of Technology. He has been an associate professor in the Department of Signal Processing and Acoustics at Aalto University, Finland, since 2016. He was a Professor at the International Audio Laboratory Erlangen, Friedrich-Alexander University, from 2013 to 2016, and a Researcher at Fraunhofer IIS from 2008 to 2013. He has contributed to several international speech and audio coding standards and is the chair and co-founder of the ISCA Special Interest Group “Security and Privacy in Speech Communication”. His research interests include technologies for spoken interaction, emphasizing efficiency and privacy, and in particular, in multi-device and multi-user environments. 
\end{IEEEbiography}

\begin{IEEEbiography}[{\includegraphics[width=1in,height=1.25in,clip,keepaspectratio]{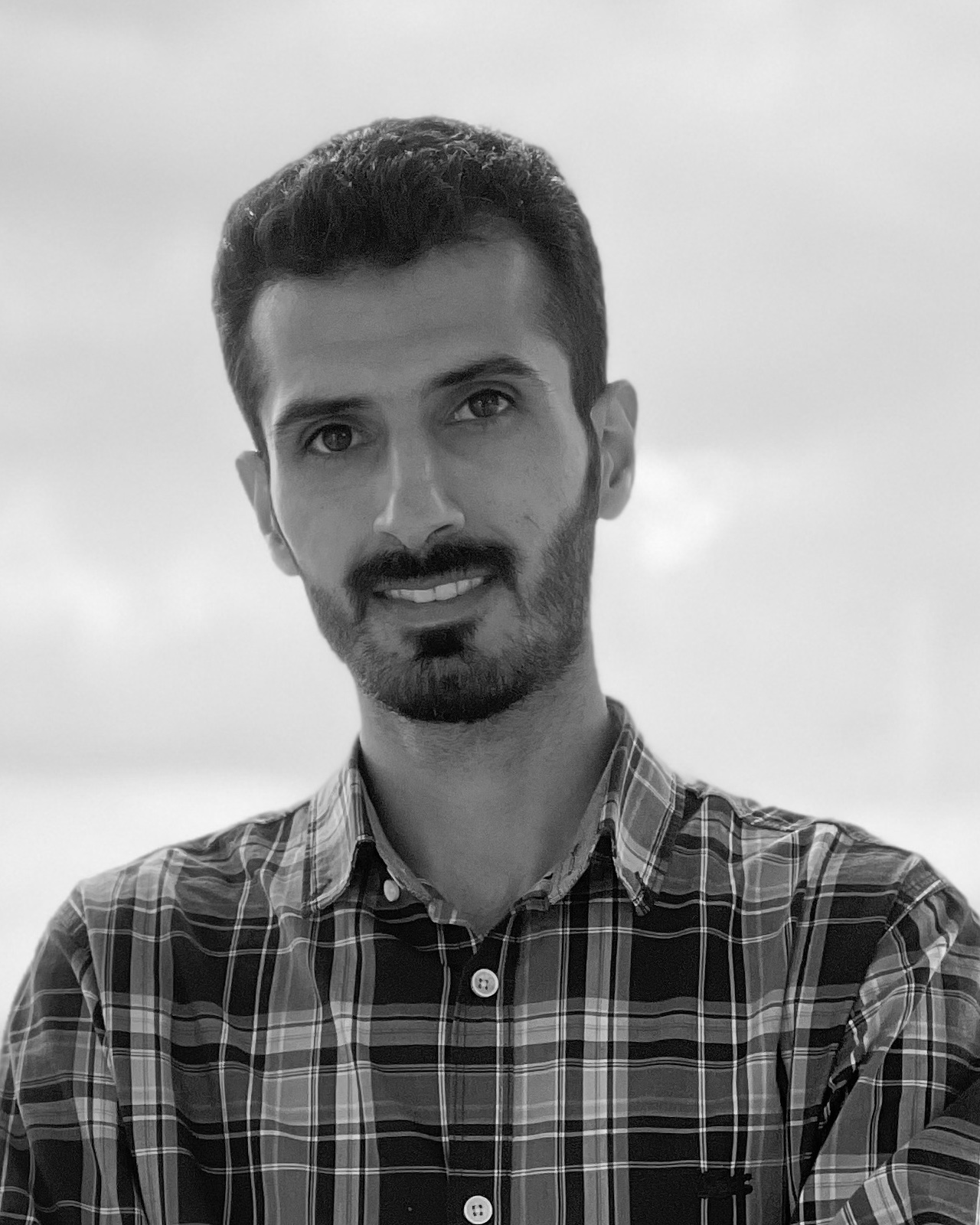}}]{Mohammad Hassan Vali} received his bachelor's and master's degrees from Babol Noshirvani University of Technology in 2014 and 2017, respectively, and his doctorate from Aalto University in 2025. He is currently a postdoctoral researcher in the Department of Computer Science at Aalto University, Finland. His research interests include image and speech processing using machine learning models, with a particular focus on vector quantization in deep neural networks for discrete representation learning.
\end{IEEEbiography}

\begin{IEEEbiography}[{\includegraphics[width=1in,height=1.25in,clip,keepaspectratio]{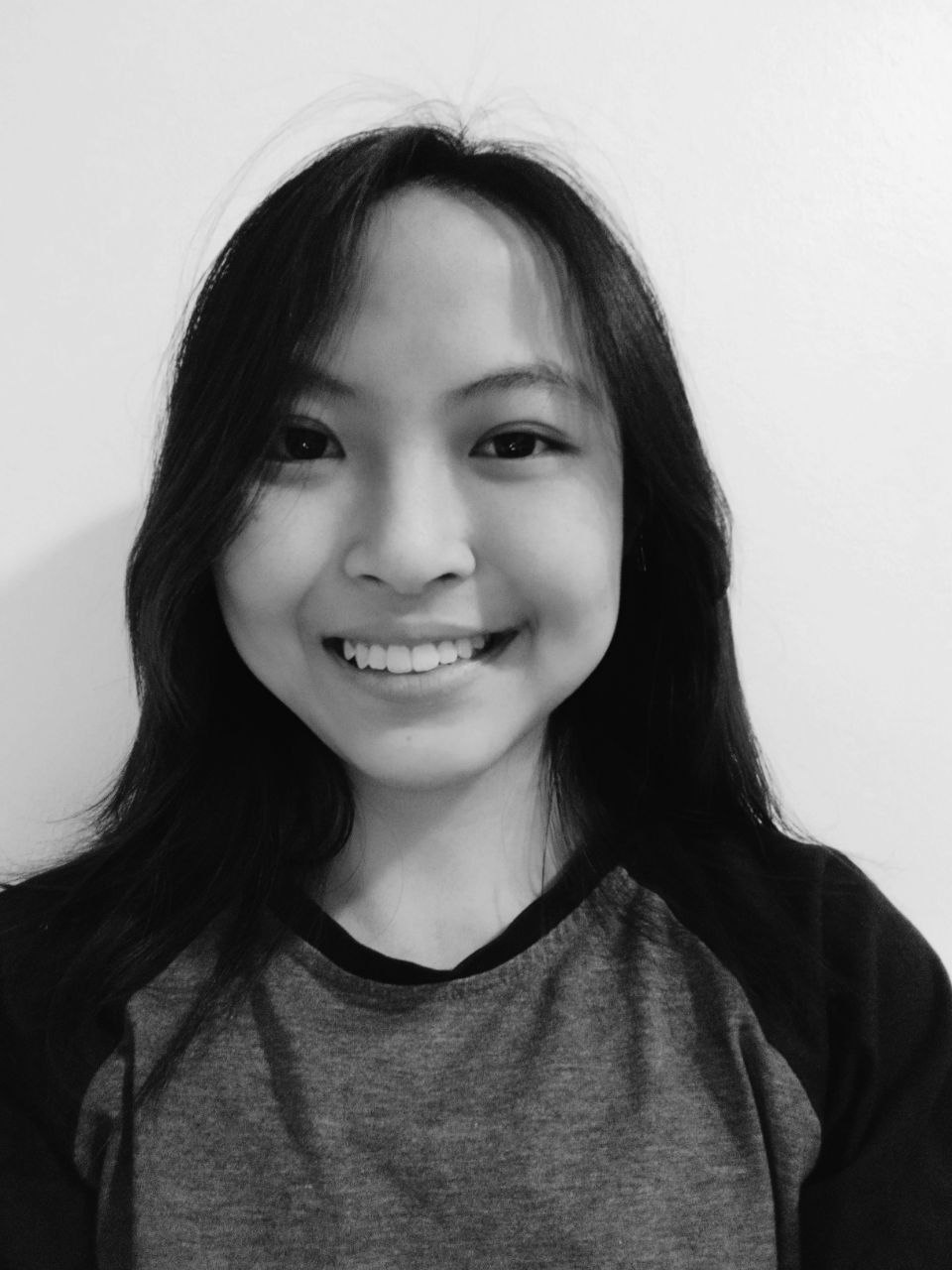}}]{My Nguyen} is a master's student in Speech and Language Technology at Aalto University, Finland. Her research interests include forensic linguistics and ethical AI. She is currently working on a master's thesis related to the application of author attribution methods in speaker identification.
\end{IEEEbiography}

\begin{IEEEbiography}[{\includegraphics[width=1in,height=1.25in,clip,keepaspectratio]{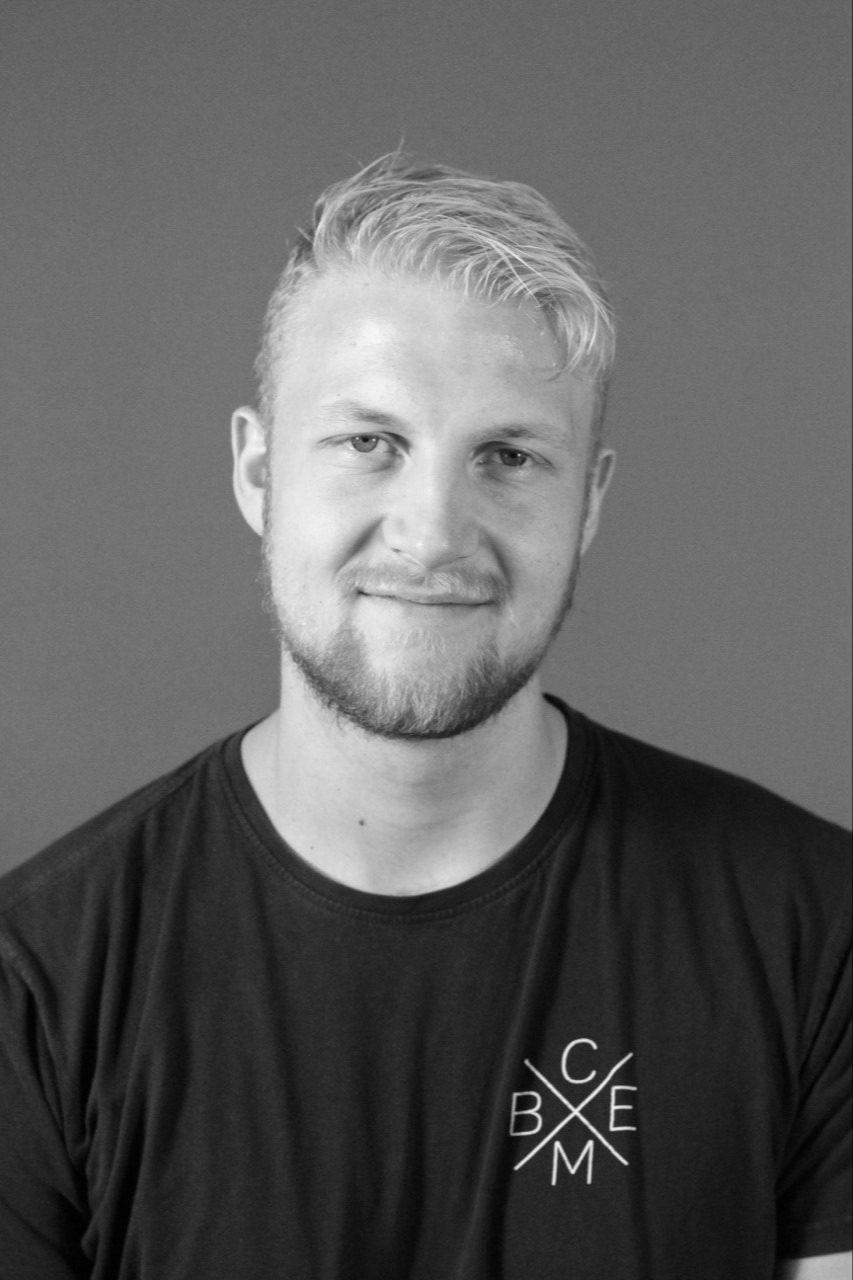}}]{Silas Rech} is a doctoral student in the Speech Interaction Technology Group at Aalto University, Finland. He received his master's degree from Aalto in 2022. His research interests focus on trustworthy, private, and ethical speech processing, in particular natural interfaces.
\end{IEEEbiography}

\end{document}